\documentclass[a4paper]{jpconf}
\usepackage{graphicx}
\usepackage{amssymb,amsmath,cite}

\newcommand{\ba}{\begin{eqnarray}}
\newcommand{\ea}{\end{eqnarray}}
\newcommand{\ban}{\begin{eqnarray*}}
\newcommand{\ean}{\end{eqnarray*}}
\newcommand{\bsub}{\begin{subequations}}
\newcommand{\esub}{\end{subequations}}
\begin{document}
\title{Exact dynamical and partial symmetries}

\author{A Leviatan}

\address{Racah Institute of Physics, The Hebrew University, 
Jerusalem 91904, Israel}

\ead{ami@phys.huji.ac.il}

\begin{abstract}
We discuss a hierarchy of broken symmetries with special emphasis on 
partial dynamical symmetries (PDS). The latter correspond to a situation 
in which a non-invariant Hamiltonian accommodates a subset of solvable 
eigenstates with good symmetry, while other eigenstates are mixed. We present 
an algorithm for constructing Hamiltonians with this property 
and demonstrate the relevance of the PDS notion 
to nuclear spectroscopy, to quantum phase transitions 
and to mixed systems with coexisting regularity and chaos. 
\end{abstract}
\section{Introduction}
\label{sec:intro}

Symmetries play an important role in dynamical systems. 
They provide quantum numbers for the classification of states, 
determine spectral degeneracies and selection rules, and facilitate 
the calculation of matrix elements.  
An exact symmetry occurs when the Hamiltonian of the system commutes 
with all the generators ($g_i$) of the symmetry-group $G$,  
$[\, \hat{H} \, , \, g_i\,] = 0$. 
In this case, all states have good symmetry and are labeled by the 
irreducible representations (irreps) of $G$. 
The Hamiltonian admits a block structure so that 
inequivalent irreps do not mix and all eigenstates 
in the same irrep are degenerate. In a dynamical symmetry 
the Hamiltonian commutes with the Casimir operator of $G$, 
$[\, \hat{H} \, , \, \hat{C}_{G}\,] = 0$, 
the block structure of $\hat{H}$ is retained, the states preserve 
the good symmetry but, in general, are no longer degenerate. 
When the symmetry is completely broken 
then $[\, \hat{H} \, , \, g_i\,] \neq 0$, and none 
of the states have good symmetry. In-between these limiting cases there 
may exist intermediate symmetry structures, called partial (dynamical) 
symmetries, for which the symmetry is neither exact nor completely broken. 
This novel concept of symmetry and its implications for 
dynamical systems are the focus of the present contribution. 

Models based on spectrum generating algebras form a convenient framework
to examine different types of symmetries and 
have been used extensively in diverse areas of physics~\cite{BNB,ibm,vibron}.
In such models the Hamiltonian is expanded in elements 
of a Lie algebra, ($G_0$), 
called the spectrum generating algebra. 
A dynamical symmetry occurs if the Hamiltonian
can be written in terms of the Casimir operators 
of a chain of nested algebras, 
$G_0\supset G_1 \supset \ldots \supset G_n$. 
The following properties are then observed. 
(i)~All states are solvable and analytic expressions
are available for energies and other observables. 
(ii)~All states are classified by quantum numbers, 
$\vert\alpha_0,\alpha_1,\ldots,\alpha_n\rangle$, 
which are the labels of the irreps of the algebras in the chain. 
(iii)~The structure of wave functions is completely dictated by symmetry
and is independent of the Hamiltonian's parameters. 
\begin{figure}[t]
\begin{center}
\includegraphics[height=15cm]{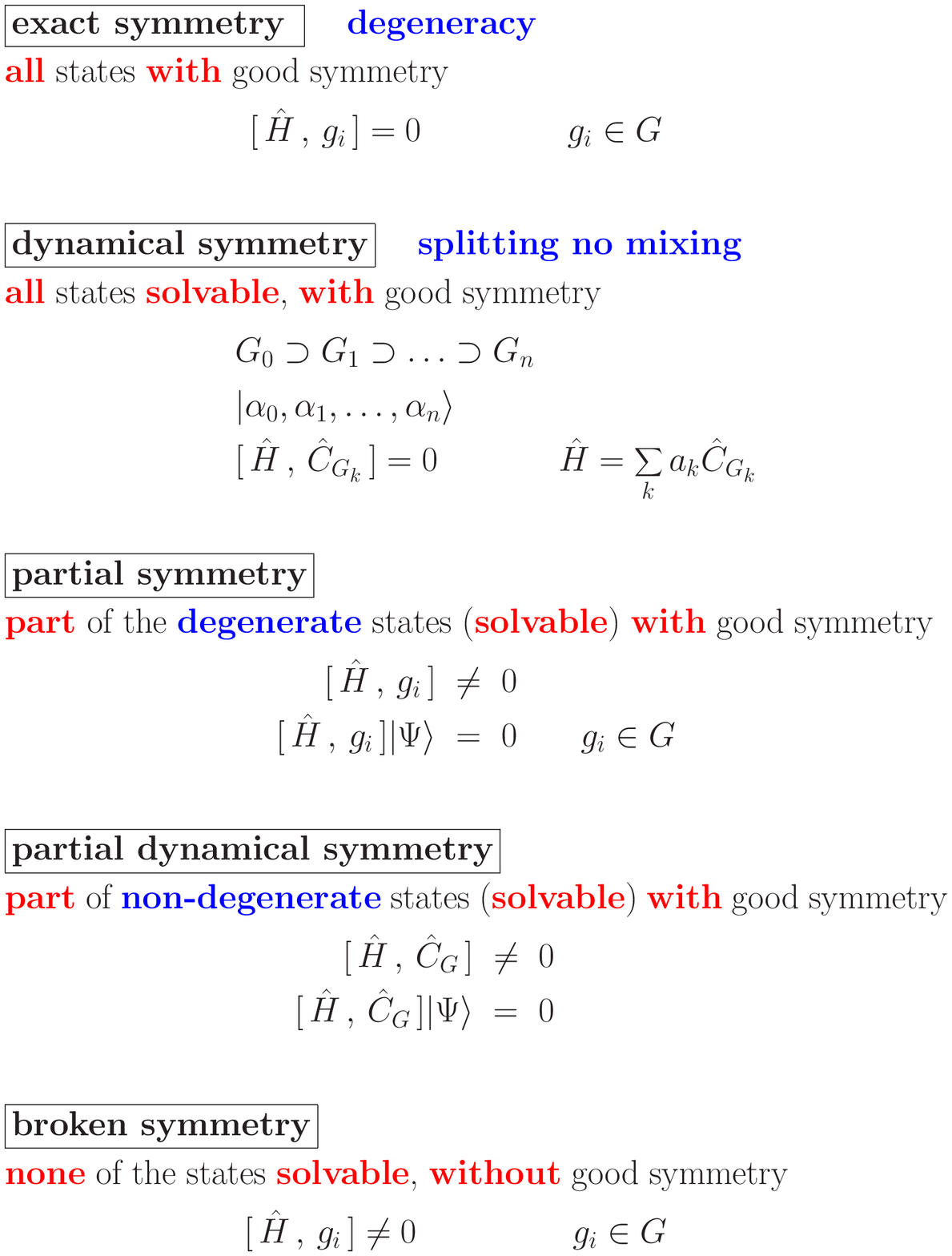}
\caption{
\protect\small 
Hierarchy of symmetries. 
\label{figSymmetry}}
\end{center}
\end{figure}

The merits of a dynamical symmetry are self-evident. 
However, in most applications to realistic systems,
the predictions of an exact dynamical symmetry are rarely fulfilled 
and one is compelled to break it. 
More often one finds that the assumed symmetry is 
not obeyed uniformly, {\it i.e.}, is fulfilled by only some states but
not by others. The required symmetry-breaking is achieved 
by including in the Hamiltonian terms associated with 
different sub-algebra chains of the parent spectrum generating algebra. 
In general, under such circumstances, solvability is lost,
there are no remaining non-trivial conserved quantum numbers and all
eigenstates are expected to be mixed.
A partial dynamical symmetry (PDS) corresponds to
a particular symmetry breaking for which some (but not all) of the 
virtues of a dynamical symmetry are retained. 
The essential idea is to relax the stringent conditions
of {\em complete} solvability
so that the properties (i)--(iii)
are only partially satisfied~\cite{lev10}. 

The notion of partial dynamical symmetry generalizes the concepts of exact and 
dynamical symmetries. In making the transition from an exact to a 
dynamical symmetry, states which 
are degenerate in the former scheme are split but not mixed in the latter,
and the block structure of the Hamiltonian is retained.
Proceeding further to partial symmetry, some blocks or selected states in a 
block remain pure, while other states mix and lose the symmetry character. 
A partial dynamical symmetry lifts 
the remaining degeneracies, but preserves the symmetry-purity of the 
selected states. The hierarchy of broken symmetries is depicted in 
Fig.~1.
 
The existence of Hamiltonians with partial symmetry or partial 
dynamical symmetry is by no means obvious. An Hamiltonian with 
the above property is not invariant under the group $G$ 
nor does it commute with the Casimir invariants of $G$, 
so that various irreps are in general mixed in its eigenstates. 
However, it posses a subset of solvable states, denoted by 
$\vert\Psi\rangle$ in Fig.~1, which respect the symmetry. 
The commutator $[\, \hat{H} \, , \, g_i\,]$ 
or $[\, \hat{H} \, , \, \hat{C}_{G}\,]$ vanishes only when it acts on 
these `special' states with good $G$-symmetry.

\section{Partial dynamical symmetries}
\label{sec:PDStypeI}

When a partial dynamical symmetry (PDS) occurs, 
the defining properties of a dynamical symmetry (DS), namely, solvability, 
good quantum numbers, and symmetry-dictated structure are fulfilled exactly, 
but by only a subset of states. 
An algorithm for constructing Hamiltonians with PDS 
has been developed in~\cite{AL92} and further elaborated 
in~\cite{RamLevVan09}. The analysis starts from the chain of nested algebras
\begin{equation}
\begin{array}{ccccccc}
G_{\rm dyn}&\supset&G&\supset&\cdots&\supset&G_{\rm sym}\\
\downarrow&&\downarrow&&&&\downarrow\\[0mm]
[h]&&\langle\Sigma\rangle&&&&\Lambda
\end{array}
\label{chain}
\end{equation}
where, below each algebra,
its associated labels of irreps are given. Eq.~(\ref{chain}) implies 
that $G_{\rm dyn}$ is the dynamical (spectrum generating) 
algebra of the 
system such that operators of all physical observables 
can be written in terms of its generators; 
a single irrep of $G_{\rm dyn}$
contains all states of relevance in the problem.
In contrast, $G_{\rm sym}$ is the symmetry algebra
and a single of its irreps contains states that are degenerate in energy. 
Assuming, for simplicity,
that particle number is conserved, then 
all states, and hence the representation $[h]$,
can then be assigned a definite particle number~$N$.
For $N$ identical particles the representation 
$[h]$ of the dynamical algebra 
$G_{\rm dyn}$ 
is either symmetric $[N]$ (bosons)
or antisymmetric $[1^N]$ (fermions)
and will be denoted, in both cases, as $[h_N]$. 
The occurrence of a DS of the type~(\ref{chain})
signifies that the Hamiltonian is written in terms of the Casimir 
operators of the algebras in the chain, 
$\hat{H}_{DS} = \sum_{G} a_{G}\,\hat{C}_{G}$, 
and the eigenstates can be labeled as
$|[h_N]\langle\Sigma\rangle\dots\Lambda\rangle$;
additional labels (indicated by $\dots$)
are suppressed in the following.
The eigenvalues of the Casimir operators in these basis states 
determine the eigenenergies $E_{DS}([h_N]\langle\Sigma\rangle\Lambda)$ 
of $\hat{H}_{DS}$. Likewise, operators can be classified
according to their tensor character under~(\ref{chain})
as $\hat T_{[h_n]\langle\sigma\rangle\lambda}$.

Of specific interest in the construction of a PDS
associated with the reduction~(\ref{chain}),
are the $n$-particle annihilation operators $\hat T$ 
which satisfy the property
\begin{equation}
\hat T_{[h_n]\langle\sigma\rangle\lambda}
|[h_N]\langle\Sigma_0\rangle\Lambda\rangle=0 ~,
\label{anni}
\end{equation}
for all possible values of $\Lambda$
contained in a given irrep~$\langle\Sigma_0\rangle$ of $G$. 
Equivalently, this condition can be phrased in terms of the action 
on a lowest weight (LW) state of the G-irrep $\langle\Sigma_0\rangle$,  
$\hat T_{[h_n]\langle\sigma\rangle\lambda}
|LW;\, [h_N]\langle\Sigma_0\rangle\rangle=0$, 
from which states of good $\Lambda$ can be obtained by projection. 
Any $n$-body, 
number-conserving normal-ordered interaction
written in terms of these annihilation operators 
and their Hermitian conjugates (which transform as the
corresponding conjugate irreps), 
$\hat{H} = \sum_{\alpha,\beta} 
A_{\alpha\beta}\, \hat{T}^{\dag}_{\alpha}\hat{T}_{\beta}$, 
has a partial G-symmetry. This comes about since for 
arbitrary coefficients, $A_{\alpha\beta}$, $\hat{H}$ 
is not a G-scalar, hence most of its eigenstates will be a 
mixture of irreps of G, yet relation~(\ref{anni}) ensures that a subset of 
its eigenstates $\vert [h_N]\langle\Sigma_0\rangle\Lambda\rangle$, 
are solvable and have good quantum numbers under the chain~(\ref{chain}). 
An Hamiltonian with partial dynamical symmetry is obtained by adding 
to $\hat{H}$ the dynamical symmetry Hamiltonian, 
$\hat{H}_{PDS} = \hat{H}_{DS} + \hat{H}$, 
still preserving the solvability
of states with $\langle\Sigma\rangle=\langle\Sigma_0\rangle$.

If the operators $\hat T_{[h_n]\langle\sigma\rangle\lambda}$ 
span the entire irrep $\langle\sigma\rangle$ of G, 
then the annihilation condition~(\ref{anni}) is satisfied
for all $\Lambda$-states in $\langle\Sigma_0\rangle$, 
if none of the $G$ irreps $\langle\Sigma\rangle$
contained in the $G_{\rm dyn}$ irrep $[h_{N-n}]$
belongs to the $G$ Kronecker product
$\langle\sigma\rangle\times\langle\Sigma_0\rangle$. 
So the problem of finding interactions
that preserve solvability
for part of the states~(\ref{chain})
is reduced to carrying out a Kronecker product. 
If relation~(\ref{anni}) holds only for some states 
$\Lambda$ in the irrep $\langle\Sigma_0\rangle$ and/or some 
components $\lambda $  of the tensor 
$\hat T_{[h_n]\langle\sigma\rangle\lambda}$, then 
the Kronecker product rule does not apply. However, 
the PDS Hamiltonian is still of the indicated normal-ordered form, 
but now the solvable states span only part of the corresponding 
$G$-irrep. The arguments for choosing the special irrep 
$\langle\Sigma\rangle=\langle\Sigma_0\rangle$ in Eq.~(\ref{anni}), 
which contains the solvable states, are based on 
physical grounds. A~frequently encountered choice is the irrep which 
contains the ground state of the system. 

In what follows we illustrate the above procedure and demonstrate the 
relevance of the PDS notion to dynamical systems. For that purpose, we 
employ the interacting boson model (IBM)~\cite{ibm}, 
widely used in the description of low-lying collective states 
in nuclei in terms of $N$ interacting monopole $(s)$ and
quadrupole $(d)$ bosons representing valence nucleon pairs.
The dynamical algebra is $G_{\rm dyn}={\rm U}(6)$
and the symmetry algebra is $G_{\rm sym}={\rm O}(3)$.
The Hamiltonian commutes with the total number operator of $s$- 
and $d$- bosons, $\hat N$, 
which is the linear Casimir of U(6). Three DS limits occur in the model
with leading subalgebras U(5), SU(3), and O(6),
corresponding to typical collective spectra observed in nuclei,
vibrational, rotational, and $\gamma$-unstable, respectively.
A geometric visualization of the model is obtained by 
an energy surface
\ba
E_{N}(\beta,\gamma) &=& 
\langle \beta,\gamma; N\vert \hat{H} \vert \beta,\gamma ; N\rangle ~, 
\label{enesurf}
\ea
defined by the expectation value of the Hamiltonian in the coherent 
(intrinsic) state~\cite{gino80,diep80}
\bsub
\ba
\vert\beta,\gamma ; N \rangle &=&
(N!)^{-1/2}(b^{\dagger}_{c})^N\,\vert 0\,\rangle ~,\\
b^{\dagger}_{c} &=& (1+\beta^2)^{-1/2}[\beta\cos\gamma 
d^{\dagger}_{0} + \beta\sin{\gamma} 
( d^{\dagger}_{2} + d^{\dagger}_{-2})/\sqrt{2} + s^{\dagger}] ~. 
\ea
\label{condgen}
\esub
Here $(\beta,\gamma)$ are
quadrupole shape parameters whose values, $(\beta_0,\gamma_0)$, 
at the global minimum of $E_{N}(\beta,\gamma)$ define the equilibrium 
shape for a given Hamiltonian. 
For a Hamiltonian with one- and two-body interactions, 
the shape can be spherical $(\beta =0)$ or 
deformed $(\beta >0)$ with $\gamma =0$ (prolate), $\gamma =\pi/3$ (oblate), 
or $\gamma$-independent. 
The equilibrium deformations associated with the DS limits are 
$\beta_0=0$ for U(5), $(\beta_0=\sqrt{2},\gamma_0=0)$ for SU(3) and 
$(\beta_0=1,\gamma_0\,{\rm arbitrary})$ for O(6). 

\section{PDS and nuclear spectroscopy}
\label{sec:su3PDStypeI}

The SU(3) DS chain of the IBM 
and related quantum numbers are given by~\cite{ibm}
\ba
\begin{array}{ccccc}
{\rm U}(6)&\supset&{\rm SU}(3)&\supset&{\rm O}(3)\\
\downarrow&&\downarrow&&\downarrow\\[0mm]
[N]&&\left (\lambda,\mu\right )& K & L
\end{array} ~,
\label{chainsu3}
\ea 
where $K$ is a multiplicity label needed for complete classification 
in the $SU(3)\supset O(3)$ reduction. 
The spectrum of the SU(3) DS Hamiltonian resembles that of an 
axially-deformed rotovibrator and the corresponding  
eigenstates are arranged in SU(3) multiplets. 
The label $K$ corresponds geometrically to the
projection of the angular momentum on the symmetry axis. 
In a given SU(3) irrep $(\lambda,\mu)$, each $K$-value is associated 
with a rotational band and states with the same angular momentum $L$, 
in different $K$-bands, are degenerate. 
The lowest SU(3) irrep is $(2N,0)$, which describes the ground band 
$g(K=0)$ of a prolate deformed nucleus. 
The first excited SU(3) irrep
$(2N-4,2)$ contains degenerate $\beta(K=0)$ and $\gamma(K=2)$ bands. 
This $\beta$-$\gamma$ degeneracy is a characteristic feature of the SU(3) 
limit which, however, is not commonly observed. 
In most deformed nuclei the $\beta$ band lies above the $\gamma$ band. 
In the IBM framework, with at most two-body interactions, one
is therefore compelled to break SU(3) 
in order to conform with the experimental data. 

The construction of Hamiltonians with SU(3)-PDS is based on identification 
of $n$-boson operators which annihilate all states in a given 
SU(3) irrep $(\lambda,\mu)$, 
chosen here to be the ground band irrep $(2N,0)$. 
For that purpose, we consider the following 
two-boson SU(3) tensors, 
$B^{\dagger}_{[n](\lambda,\mu)K;L m}$, with $n=2$, 
$(\lambda,\mu)=(0,2)$ and 
angular momentum $L =0,\,2$
\bsub
\ba
B^{\dagger}_{[2](0,2)0;00} &\propto& 
P^{\dagger}_{0} = d^{\dagger}\cdot d^{\dagger} - 2(s^{\dagger})^2 ~,\\
B^{\dagger}_{[2](0,2)0;2m} &\propto& 
P^{\dagger}_{2m} = 2d^{\dagger}_{m}s^{\dagger} + 
\sqrt{7}\, (d^{\dagger}\,d^{\dagger})^{(2)}_{m} ~.
\ea
\label{PL}
\esub 
The corresponding Hermitian conjugate boson-pair annihilation operators,  
$P_0$ and $P_{2m}$, transform 
as $(2,0)$ under SU(3), and satisfy
\ba
P_{0}\,\vert [N](2N,0)K=0, L\rangle &=& 0 ~,
\nonumber\\
P_{2m}\,\vert [N](2N,0)K=0, L\rangle &=& 0 ~.
\label{P0P2}
\ea
The indicated $L$-states span the entire SU(3) irrep 
$(\lambda,\mu)=(2N,0)$. 
They can be obtained by angular momentum projection from 
the coherent state, $\vert\beta=\sqrt{2},\gamma=0 ; N \rangle$, 
of Eq.~(\ref{condgen}), which is the 
lowest-weight state of this irrep 
and serves as an intrinsic state for the SU(3) ground band. 
The relations in Eq.~(\ref{P0P2}) follow from the 
fact that the action of the operators $P_{Lm}$ leads to a state with 
$N-2$ bosons in the U(6) irrep $[N-2]$, 
which does not contain the SU(3) irreps obtained from the product 
$(2,0)\times (2N,0)= (2N+2,0)\oplus (2N,1)\oplus (2N-2,2)$. 
In addition,  $P_0$ satisfies 
\ba
P_{0}\,\vert [N](2N-4k,2k)K=2k, LM \rangle &=& 0 ~, 
\label{P0}
\ea
where for $k> 0$ the indicated $L$-states 
span only part of the SU(3) irreps 
$(\lambda,\mu)=(2N-4k,2k)$ and form the 
rotational members of excited 
$\gamma^{k}(K=2k)$ bands. 

Following the general algorithm, a two-body Hamiltonian with partial 
SU(3) symmetry can now be constructed as~\cite{AL92,lev96}
\ba
\hat{H}(h_0,h_2) &=& h_{0}\, P^{\dagger}_{0}P_{0} 
+ h_{2}\,P^{\dagger}_{2}\cdot \tilde{P}_{2} ~,
\label{HPSsu3}
\ea
where $\tilde P_{2m} = (-)^{m}P_{2,-m}$ and the dot denotes a scalar 
product. For $h_{2}=h_{0}$, the Hamiltonian is an SU(3) scalar, 
related to the quadratic Casimir operator of SU(3), 
for $h_0=-5h_2$ it transforms as a $(2,2)$ SU(3) 
tensor component and has the form 
\ba
\hat{H}(h_0,h_2) &=& 
h_{2}\left [-\hat C_{{\rm SU(3)}} + 2\hat N (2\hat N+3)\right ]
+ (h_0 - h_2)\, P^{\dagger}_{0}P_{0} ~.\quad
\label{HPS2su3}
\ea 
The first term in Eq.~(\ref{HPS2su3}) belongs to the SU(3) DS Hamiltonian. 
The $P^{\dag}_{0}P_0$ term is not diagonal in the SU(3) chain, however,  
Eqs.~(\ref{P0P2})-(\ref{P0}) ensure that 
$\hat{H}(h_0,h_2)$ retains selected solvable states with good 
SU(3) symmetry. Specifically, the solvable states are members of the 
ground $g(K=0)$ and $\gamma^{k}(K=2k)$ bands with the following 
characteristics
\ba
\begin{array}{lll}
\vert N,(2N,0)K=0,L\rangle & E=0 & L=0,2,4,\ldots, 2N\\[7pt]
\vert N,(2N-4k,2k)K=2k,L\rangle & 
E =  h_{2}\,6k \left (2N - 2k+1 \right ) &
L=K,K+1,\ldots, (2N-2k)~.
\end{array}\;\;
\label{ggamband}
\ea
\begin{figure}[t]
\begin{center}
\includegraphics[height=8cm]{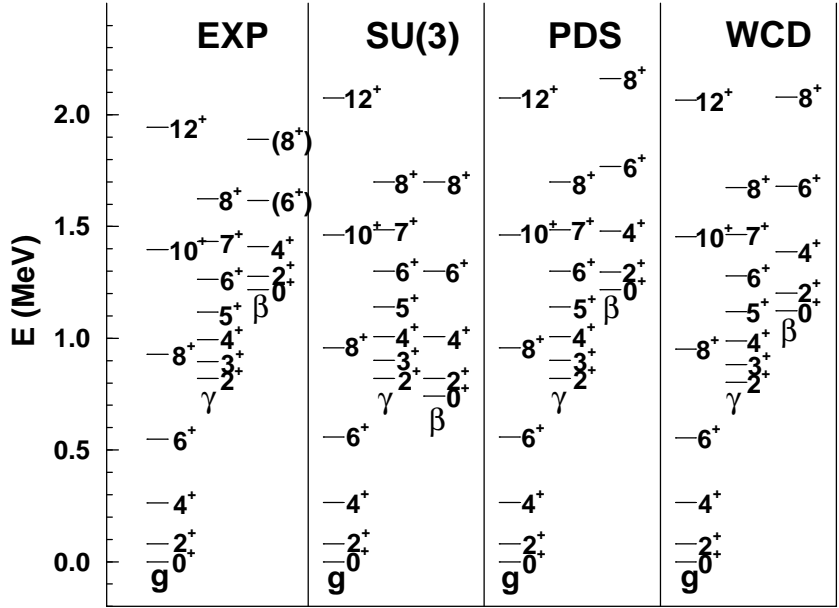}
\caption{
\small
Spectra of $^{168}$Er ($N=16$). Experimental energies
(EXP) are compared with IBM calculations in an exact SU(3) dynamical 
symmetry [SU(3)], in a broken SU(3) symmetry (WCD) 
and in a partial dynamical SU(3) symmetry (PDS). 
The latter employs the Hamiltonian of Eq.~(\ref{HPSsu3}), 
$\hat{H}(h_0,h_2) + C\,\hat{C}_{{\rm O(3)}}$, with 
$h_0=8,\,h_2=4,\,C=13$ keV. 
\label{figEr168}}
\end{center}
\end{figure}
The remaining eigenstates of $\hat{H}(h_0,h_2)$ do 
not preserve the SU(3) symmetry and therefore get mixed. 
This situation corresponds precisely to that of partial SU(3) symmetry. 
An Hamiltonian $\hat{H}(h_0,h_2)$ which is not an SU(3) scalar has a 
subset of {\it solvable} eigenstates which continue to have 
good SU(3) symmetry. One can add the Casimir operator of O(3), 
$\hat{C}_{{\rm O(3)}}$, to $\hat{H}(h_0,h_2)$ and by doing so, 
convert the partial SU(3) symmetry into partial dynamical SU(3) symmetry. 
The additional rotational term contributes just an $L(L+1)$ splitting 
but does not affect the wave functions.

The empirical spectrum of $^{168}$Er is shown in 
Fig.~\ref{figEr168} and compared with SU(3)-DS, SU(3)-PDS and broken 
SU(3) calculations~\cite{lev96}. The SU(3)-PDS spectrum shows an
improvement over the schematic, exact SU(3) dynamical symmetry 
description, since the $\beta$-$\gamma$ degeneracy is lifted. 
The quality of the calculated PDS spectrum is similar to that obtained
in the broken-SU(3) calculation, however, in the former 
the ground $g(K=0_1)$ and $\gamma(K=2_1)$ bands remain solvable 
with good SU(3) symmetry, $(\lambda,\mu)=(2N,0)$ and $(2N-4,2)$ respectively. 
At the same time, the excited $K=0^{+}_2$ band involves about $13\%$ 
SU(3) admixtures into the dominant $(2N-4,2)$ irrep. 
Since the wave functions of the solvable 
states~(\ref{ggamband}) are known, one can obtain 
{\it analytic} expressions for matrix elements of observables between them. 
In particular, the calculated B(E2) ratios 
for $\gamma\to g$ transitions lead to parameter-free predictions 
in excellent agreement with experiment~\cite{lev96}, thus 
confirming the relevance of SU(3)-PDS to the spectroscopy 
of $^{168}$Er.

\section{PDS and quantum phase transitions}
\label{sec:PDSQPT}

Quantum phase transitions (QPT) 
occur at zero temperature as a function of a 
coupling constant in the Hamiltonian. Such ground-state energy phase 
transitions are a pervasive 
phenomenon observed in many branches of physics, and are realized 
empirically in nuclei as transitions between different shapes. 
QPTs occur as a result of a competition between terms in the Hamiltonian 
with different symmetry character, which lead to considerable mixing in the 
eigenfunctions, especially at the critical-point where the structure 
changes most rapidly. An interesting question to address is 
whether there are any symmetries (or traces of) still present at the 
critical points of QPT. As shown below, unexpectedly, partial dynamical 
symmetries can survive at the critical point in spite of the 
strong mixing~\cite{lev07}. 
\begin{figure}[t]
\begin{minipage}{18pc}
\includegraphics[width=2.1in,angle=270,clip=]
{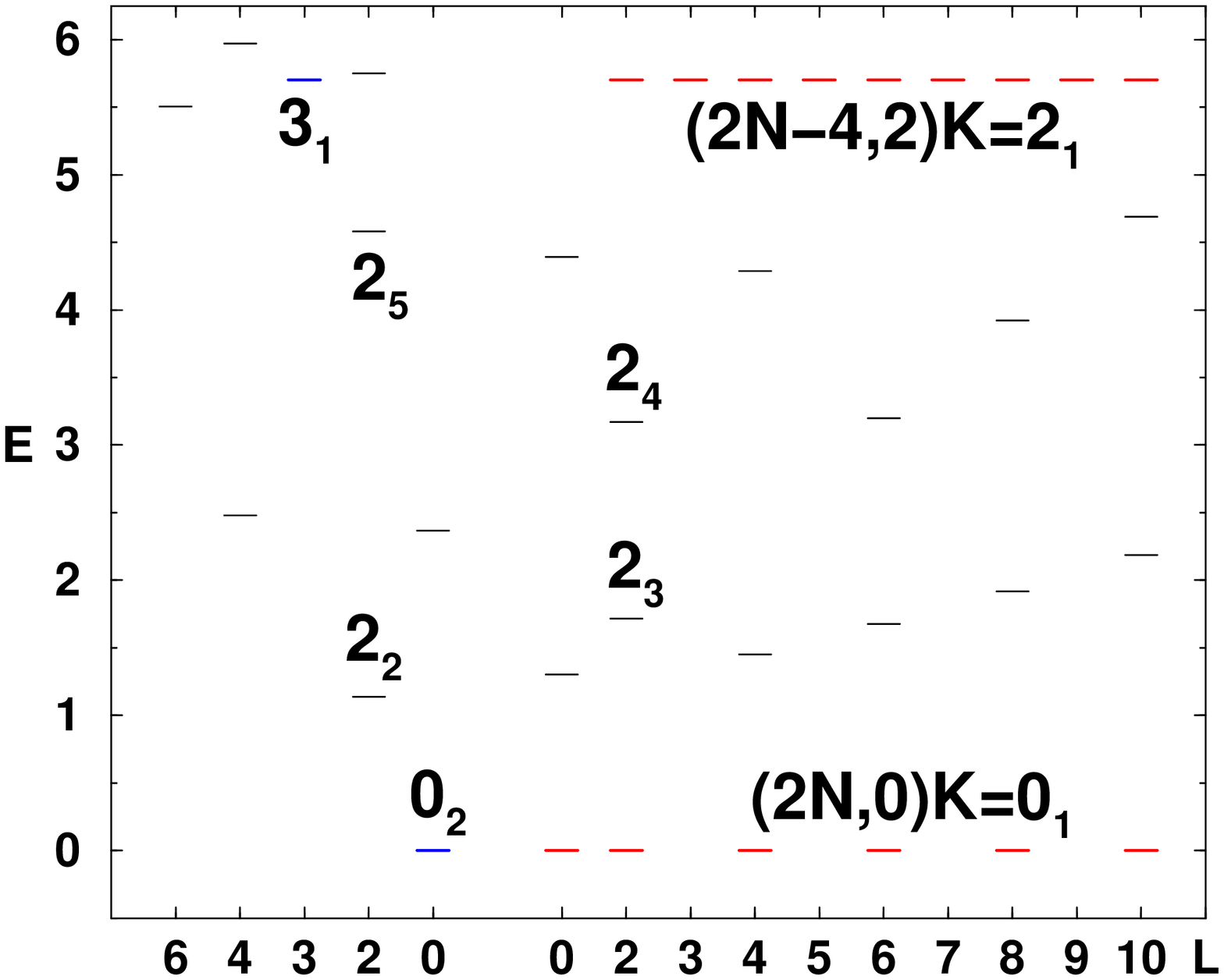}
\caption{\label{fig3}
\small
Spectrum of $\hat{H}_{cri}$, 
Eq.~(\ref{hcri1st}), with $h_2=0.05$ and $N=10$. 
$L(K=0_1)$ and $L(K=2_1)$ are the solvable SU(3) states 
of Eq.~(\ref{ggamband}) with $k=0$ and $k=1$, respectively. 
$L=0_2,3_1$ are the solvable U(5) states of Eq.~(\ref{spher}).}
\end{minipage}\hspace{0.7cm}%
\begin{minipage}{17pc}
\includegraphics[width=2.1in,angle=270,clip=]
{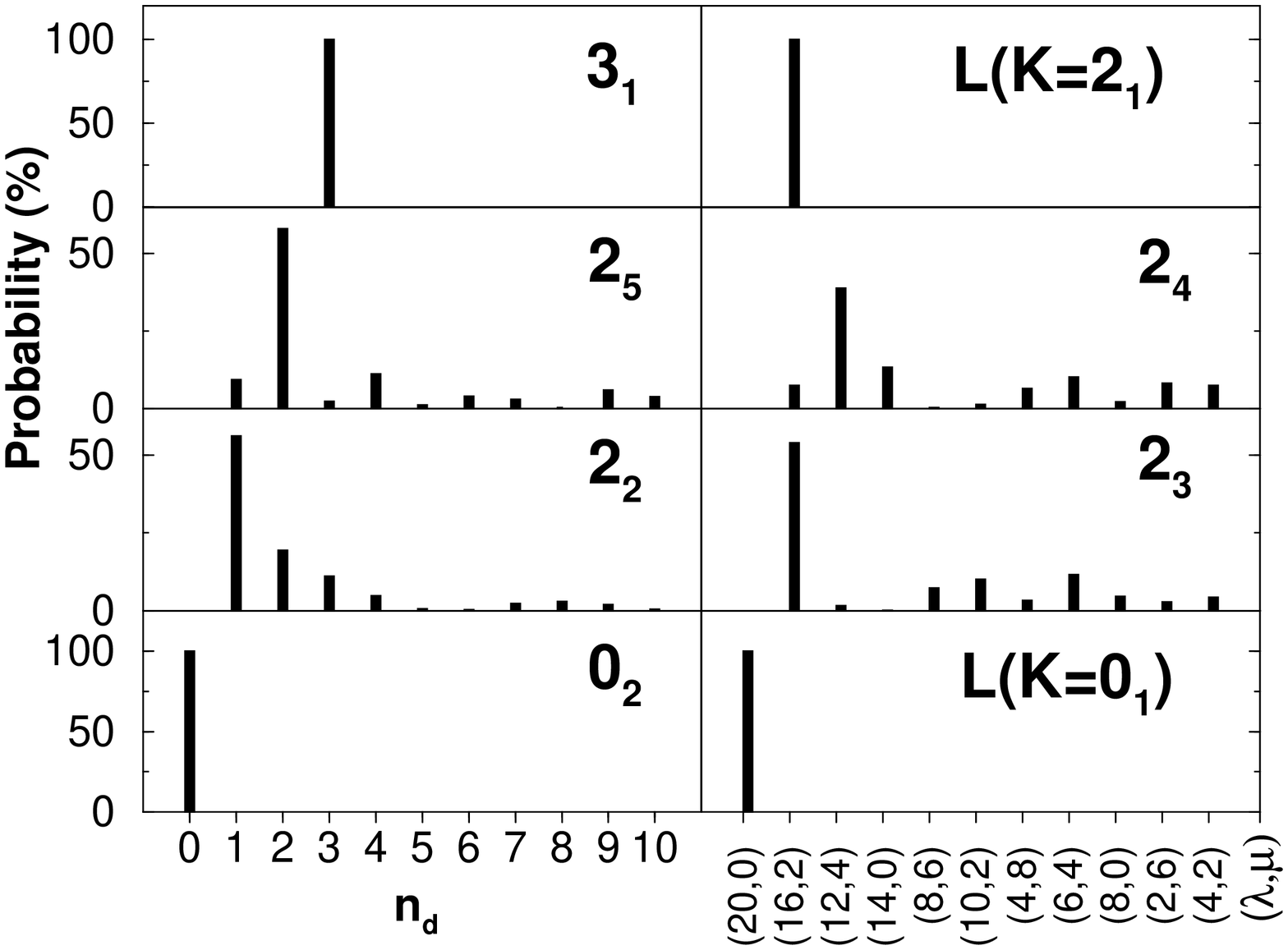}
\caption{\label{fig4}
\small
U(5) ($n_d$) and SU(3) $[(\lambda,\mu)]$ decomposition for 
selected spherical and deformed states in Fig.~\ref{fig3}.}
\end{minipage} 
\end{figure}

A convenient framework to study 
symmetry aspects of QPTs in nuclei is the IBM~~\cite{ibm}, whose 
dynamical symmetries 
correspond to possible phases of the system. The relevant Hamiltonian 
in such a study involves terms with from different DS chains. 
The nature of the phase transition is 
governed by the topology of the corresponding surface~(\ref{enesurf}), 
which serves as a Landau's potential 
with the equilibrium deformations as order parameters. 
The surface at the critical-point of a first-order transition 
is required to have two-degenerate minima, corresponding to the two 
coexisting phases. For example, the following first-order critical surface 
\ba
E_{N}(\beta,\gamma=0) &=& 
2h_{2}N(N-1)(1+\beta^2)^{-2}\beta^2\left ( \beta-\sqrt{2}\,\right )^2 ~,
\ea 
has degenerate spherical and 
deformed minima at $\beta=0$ and $(\beta=\sqrt{2},\gamma=0)$, 
corresponding to spherical 
and axially-deformed shapes. 
A barrier of height $h = 2h_{2}N(N-1)(1-\sqrt{3}/2)$ 
separates the two minima. 
Such a surface can be obtained from Eq.~(\ref{enesurf}) with 
the following Hamiltonian
\ba
\hat{H}_{cri} &=& h_{2}\, 
P^{\dagger}_{2}\cdot\tilde{P}_{2} ~, 
\label{hcri1st}
\ea 
which is qualified to be a critical Hamiltonian of a first-order transition 
between these shapes. 
$\hat{H}_{cri}$ is recognized to be a special case of the Hamiltonian 
of Eq.~(\ref{HPSsu3}), shown to have SU(3)-PDS. 
As such, it has a subset of solvable eigenstates, 
Eq.~(\ref{ggamband}), 
which are members of deformed ground 
$g(K=0)$ and $\gamma^{k}(K=2k)$ bands 
with good SU(3) symmetry, $(\lambda,\mu)=(2N-4k,2k)$. 
In addition, $H_{cri}$ has also the following 
solvable spherical eigenstates with good U(5) symmetry 
\bsub
\ba 
\vert N,n_d=\tau=L=0 \rangle  \;\; &&E = 0 ~,
\label{nd0b0}\\
\vert N,n_d=\tau=L=3 \rangle \;\;
&&E = 6 h_2(2N-1)~.
\qquad\quad
\label{nd3b0}
\ea
\label{spher}
\esub
These are selected basis states, $\vert[N],n_d,\tau,L\rangle$, of the 
U(5)-DS chain $U(6)\supset U(5)\supset O(5)\supset O(3)$. 
$H_{cri}$ (\ref{hcri1st}) is not invariant under $U(5)$, nor does it 
have a U(5)-DS, yet it has solvable states with good U(5) symmetry. 
By definition, it posses a U(5)-PDS. 
The spherical $L=0$ state, Eq.~(\ref{nd0b0}), is 
exactly degenerate with the SU(3) ground band, Eq.~(\ref{ggamband}) with 
$k=0$, and the spherical $L=3$ state, Eq.~(\ref{nd3b0}), 
is degenerate with the SU(3) $\gamma$-band, 
Eq.~(\ref{ggamband}) with $k=1$. 
The remaining levels of $\hat{H}_{cri}$, shown 
in Fig.~\ref{fig3} are calculated numerically. 
Their wave functions are spread over many 
U(5) and SU(3) irreps, as is evident from 
the $n_d$ and $(\lambda,\mu)$ decomposition shown in Fig.~\ref{fig4}. 
This situation, 
where some states are solvable with good U(5) symmetry, 
some are solvable with good SU(3) symmetry and all other 
states are mixed with respect to both U(5) and SU(3), 
defines a U(5) PDS coexisting with a SU(3) PDS. 
The presence in the spectrum of both spherical states 
(dominated by a single $n_d$ 
component) and deformed states arranged in bands (with a broad $n_d$ 
distribution) signals a first-order transition. 

The above results demonstrate the relevance of the PDS notion 
to critical-points of QPT, with phases characterized by Lie-algebraic 
symmetries. In the example considered, first-order critical Hamiltonians 
exhibit distinct subsets of solvable states with good symmetries, 
giving rise to a coexistence of different PDS. 
The ingredients of an algebraic description 
of QPT is a spectrum generating algebra and an associated geometric 
space, formulated in terms of coherent (intrinsic) states. 
The same ingredients are used in the construction of Hamiltonians 
with PDS. These, in accord with the present discussion, 
can be used as tools to explore the role of partial symmetries in 
governing the critical behaviour of dynamical systems undergoing QPT.

\section{PDS and mixed regular and chaotic dynamics}
\label{sec:PDSChaos}

Partial dynamical symmetries can play a role not only for discrete 
spectroscopy but also for analyzing statistical aspects of 
nonintegrable systems. 
Hamiltonians with a dynamical symmetry are always completely integrable. 
The Casimir invariants of the algebras in the chain provide a set 
of constants of the motion in involution. The classical motion is purely 
regular. A symmetry-breaking is connected 
to nonintegrability and may give rise to chaotic motion. 
Hamiltonians with PDS are not completely integrable, 
hence can exhibit stochastic behavior, nor are they completely chaotic, 
since some eigenstates preserve the symmetry exactly. 
Consequently, such Hamiltonians are optimally suitable to the study
of mixed systems with coexisting regularity and chaos. 

The dynamics of a generic classical Hamiltonian system is mixed;
KAM islands of regular motion and chaotic regions coexist
in phase space. 
In the associated quantum system, if no separation between regular and 
irregular states is done, the statistical properties of the spectrum 
are usually intermediate between the Poisson and the Gaussian orthogonal 
ensemble (GOE) statistics. 
In a PDS, the symmetry of the subset of solvable states 
is exact, yet does not arise from invariance properties of the 
Hamiltonian. If the fraction of solvable states 
remains finite in the classical limit, one might expect that a corresponding 
fraction of the phase space would consist of KAM tori and 
exhibit regular motion. It turns out that 
PDS has an even greater effect on the dynamics. 
It is strongly 
correlated with suppression ({\it i.e.}, reduction) of chaos even though the 
fraction of solvable states approaches zero in the classical 
limit~\cite{walev93,levwhe96}. 

We consider the following IBM Hamiltonian
\ba
\hat{H}(\beta_0) &=& h_{0}\, 
P^{\dagger}_{0}(\beta_{0})P_{0}(\beta_0) 
+ h_{2}\,P^{\dagger}_{2}(\beta_0)\cdot \tilde{P}_{2}(\beta_0) ~, 
\label{Hchaos}
\ea 
where $P^{\dagger}_{0}(\beta_0) = 
d^{\dagger}\cdot d^{\dagger} - \beta_{0}^2(s^{\dagger})^2$ and 
$P^{\dagger}_{2m}(\beta_0) = 
\beta_{0}\sqrt{2}\,d^{\dagger}_{m}s^{\dagger} + 
\sqrt{7}\, (d^{\dagger}\,d^{\dagger})^{(2)}_{m}$.
For $\beta_0=\sqrt{2}$, $\hat{H}(\beta_0=\sqrt{2})$ reduces to the 
Hamiltonian of Eq.~(\ref{HPSsu3}), which has  SU(3)-PDS with a subset 
of solvable states listed in Eq.~(\ref{ggamband}). 
At a given spin per boson $l=L/N$, and 
to leading order in $1/N$, the fraction $f$ of solvable states 
decreases like $1/N^2$ with boson number.
However, at a given boson number $N$, this fraction increases with $l$, 
a feature which is valid also for finite $N$~\cite{walev93}. 
The classical Hamiltonian is obtained from~(\ref{Hchaos}) 
by replacing operators by c-numbers, 
$s^{\dag},d^{\dag}_{\mu}\to\alpha_{s}^{*},\alpha_{\mu}^{*}$ 
and taking $N\to\infty$, with $1/N$ playing the role of $\hbar$. 
\begin{figure}[t]
\begin{center} 
\includegraphics[height=5in]{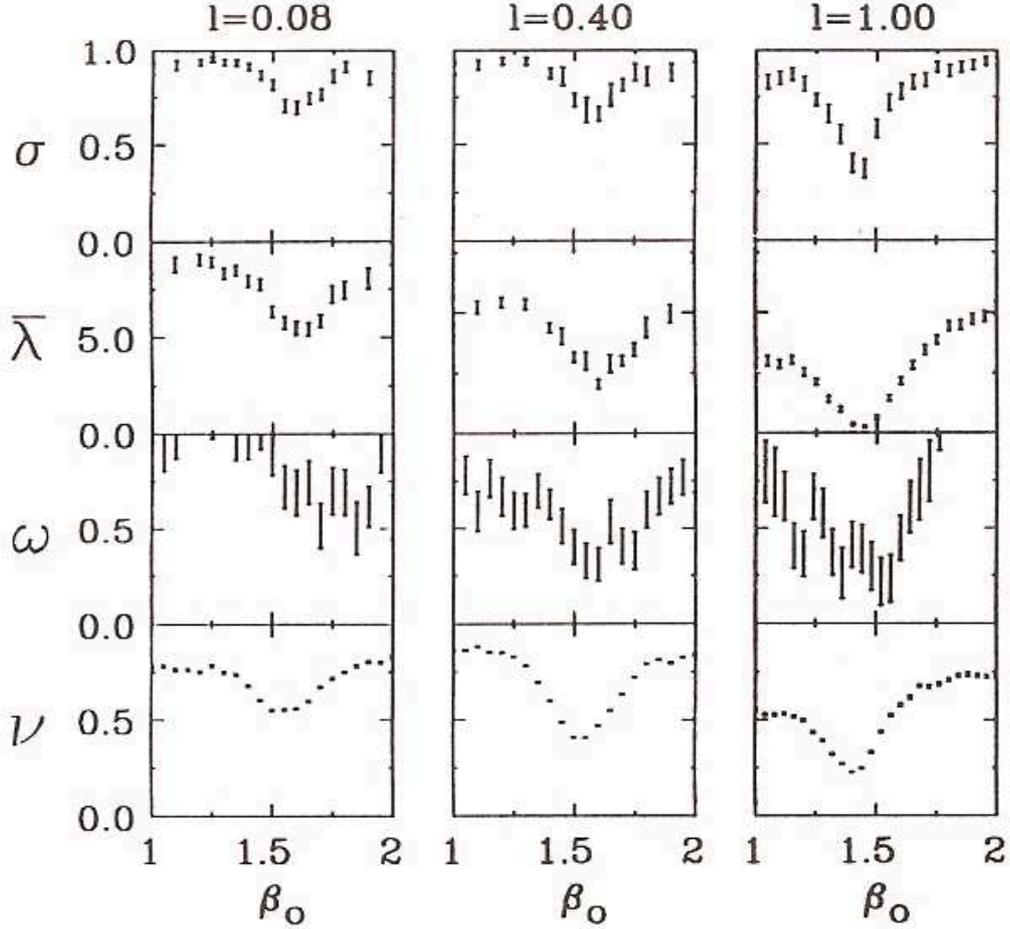}
\caption{
\small
Classical $(\sigma,\bar{\lambda})$ and quantal $(\omega,\nu)$ measures of 
chaos versus $\beta_0$ for the Hamiltonian~(\ref{Hchaos}) with $h_2/h_0=7.5$. 
Shown are three cases with classical spins $l=0.08,\, 0.4$, and $1$. 
The quantal calculations $(\omega,\nu)$ are done for $N=25$ bosons and spins 
$L=2,10$, and $25$, respectively. Notice that with increasing spin the 
minimum gets deeper and closer to $\beta_0=\sqrt{2}$. The suppression of 
chaos near $\beta_0=\sqrt{2}$ is seen both for finite $N$ through the 
measures $\omega,\,\nu$ and in the classical limit $N\to\infty$ through the 
measures $\sigma,\,\bar{\lambda}$. 
\label{figchaossu3}}
\end{center}
\end{figure}

To study the effect of the SU(3) PDS on the dynamics, 
we fix the ratio $h_{2}/h_{0}$ at a value far from the exact SU(3) symmetry 
(for which $h_0/h_2 =1)$. 
We then change $\beta_0$ in the range $1\leq\beta_0\leq 2$. 
Classically, we determine the fraction $\sigma$ of chaotic volume and the 
average largest Lyapunov exponent $\bar{\lambda}$. To analyze 
the quantum Hamiltonian, we study spectral and transition intensity 
distributions. The nearest neighbors level spacing distribution is 
fitted by a Brody distribution, 
$P_{\omega}(S) = AS^{\omega}\exp(-\alpha S^{1+\omega})$, 
where $A$ and $\alpha$ are determined by 
the conditions that $P_{\omega}(S)$ is normalized to $1$ and 
$\langle S\rangle =1$. For the Poisson statistics $\omega=0$ and for 
GOE $\omega=1$, corresponding to integrable and fully chaotic classical 
motion, respectively. The intensity distribution 
of the E2 operator 
is fitted by a $\chi^2$ distribution in $\nu$ degrees of 
freedom, $P_{\nu}(y) = 
[(\nu/2\langle y\rangle)^{\nu/2}/\Gamma(\nu/2)]y^{\nu/2-1}
\exp(-\nu y/2\langle y\rangle)$. 
For the GOE, $\nu=1$ and $\nu$ decreases as the dynamics become more 
regular.

Fig.~\ref{figchaossu3} 
shows the two classical measures $\sigma$, $\bar{\lambda}$ and 
the two quantum measures $\omega$, $\nu$ for the Hamiltonian~(\ref{Hchaos}) 
as a function of $\beta_0$. The parameters of the Hamiltonian are taken to be 
$h_2/h_0=7.5$ and the number of bosons is $N=25$. Shown are three 
classical spins $l=0.08,\, 0.4$ and $1$, which correspond in the 
quantum case to $L=2,\, 10$ and $25$. All measures show a pronounced minimum 
which gets deeper and closer to $\beta_0=\sqrt{2}$ [where the partial SU(3) 
symmetry occurs] as the classical spin increases. This behaviour is 
correlated with the fraction of solvable states (at a constant $N$) being 
larger at higher $l$. 
We remark that the classical measures show a clear enhancement of the 
regular motion near $\beta_0=\sqrt{2}$ even though the fraction of solvable 
states vanishes as $1/N^2$ in the classical limit $N\to\infty$. 
That the observed suppression of chaos is related to the SU(3) 
PDS, is confirmed by the fact that the SU(3) entropy averaged over all 
eigenstates of $\hat{H}(\beta_0)$ displays a minimum which is well 
correlated with the minimum in Fig.~\ref{figchaossu3}. 
The existence of an SU(3) PDS 
seems to have an effect of increasing the SU(3) symmetry 
of all states, not just those with an exact SU(3) symmetry. 

The following physical picture emerges from the 
analysis of low-dimensional systems~\cite{walev93,levwhe96}. 
At the quantum level, PDS 
by definition implies the existence
of a ``special'' subset of  states, which observe the symmetry. The PDS 
affects the purity
of other states in the system; in particular, neighboring states,
accessible by perturbation theory,
possess approximately good symmetry. Analogously,
at the classical level, the region of phase space near the ``special''
torus also has toroidal structure. As a consequence
of having PDS, 
a finite region of phase space is
regular and a finite fraction of states is approximately ``special''.
This clarifies the observed suppression of chaos. 

\section{Concluding remarks}
\label{conclusion}

Underlying the PDS notion, is the recognition that 
a non-invariant Hamiltonian can have selected eigenstates 
with good symmetry and good quantum numbers. 
In such a case, the symmetry in question is preserved 
in some states but is broken in the Hamiltonian (an opposite situation to 
that encountered in a spontaneously-broken symmetry). 
PDSs appear to be a common feature in algebraic descriptions of dynamical 
systems. They are not restricted to a specific model but can be applied 
to any quantal systems of interacting particles, bosons and 
fermions~\cite{lev10,escher00,isa08}. 

In PDS of type~I described above, only part of the 
eigenspectrum is analytically solvable and retains all the dynamical 
symmetry (DS) quantum numbers. Additional types of PDS are possible. 
In PDS of type~II, the 
entire eigenspectrum retains some of the DS quantum numbers~\cite{isa99}. 
PDS of type~III has a hybrid character, in the sense that 
some (solvable) eigenstates keep some of the quantum 
numbers~\cite{levisa02}. General algorithms for selecting and 
constructing Hamiltonians with PDSs of various types are available. 
The advantage of using interactions with a PDS is that they can be introduced, 
in a controlled manner, without destroying results previously obtained 
with a DS for a segment of the spectrum. These virtues 
generate an efficient tool which 
greatly enhance the scope of applications of algebraic modeling 
of quantum many-body systems. 
\ack
This work is supported by the Israel Science Foundation and the BSF.

\end{document}